# Superconductivity in $La_{1-x}Ce_xOBiSSe$: carrier doping by mixed valence of Ce ions

Ryota Sogabe[1], Yosuke Goto[1(a)], Atsuhiro Nishida[1], Takayoshi Katase[2,3], and Yoshikazu Mizuguchi[1]

[1] *Department of Physics, Tokyo Metropolitan University, Hachioji 192-0397, Japan*
[2] *Laboratory for Materials and Structures, Institute of Innovative Research, Tokyo Institute of Technology, 4259 Nagatsuta, Midori, Yokohama, 226−8503, Japan*
[3] *PRESTO, Japan Science and Technology Agency, 7 Gobancho, Chiyoda, Tokyo, 102-0076, Japan*

**Abstract** – We report the effects of Ce substitution on structural, electronic, and magnetic properties of layered bismuth-chalcogenide $La_{1-x}Ce_xOBiSSe$ ($x$ = 0–0.9), which are newly obtained in this study. Metallic conductivity was observed for $x \geq 0.1$ because of electron carriers induced by mixed valence of Ce ions, as revealed by bond valence sum calculation and magnetization measurements. Zero resistivity and clear diamagnetic susceptibility were obtained for $x$ = 0.2–0.6, indicating the emergence of bulk superconductivity in these compounds. Dome-shaped superconductivity phase diagram with the highest transition temperature ($T_c$) of 3.1 K, which is slightly lower than that of F-doped LaOBiSSe ($T_c$ = 3.7 K), was established. The present study clearly shows the mixed valence of Ce ions can be utilized as an alternative approach for electron-doping in layered bismuth-chalcogenides to induce superconductivity.

**Introduction** – Since the discovery of $BiCh_2$-based (Ch: S, Se) superconductors in 2012, this family of compounds has received much attention as a new class of layered superconductors [1–3]. The crystal structure composed of alternate stacks of electrically conducting $BiCh_2$ layers and insulating (blocking) layers, which is schematically depicted in the inset of Fig. 1, is similar to that of cuprate or Fe-based high-transition-temperature ($T_c$) superconductors [4,5]. Several types of $BiCh_2$-based superconductors have been reported, and the record $T_c$ of 11 K was obtained so far [3,6]. Although many experimental and theoretical studies have been conducted to clarify the superconductivity mechanism and to increase the $T_c$ of $BiCh_2$-based superconductors, the nature of superconductivity is still under debate [7]. In the early studies including theoretical calculation [8], Raman scattering [9], muon-spin spectroscopy [10], and thermal conductivity [11] experiments suggested conventional mechanisms with fully gapped s-wave. However, recent first-principles calculation [12], angle-resolved photoemission spectroscopy [13], and Se isotope effect [14] proposed unconventional pairing mechanism in $BiCh_2$-based superconductors. Therefore, systematic characterization is still crucial to clarify the superconductivity mechanisms of these compounds.

Primarily, superconductivity in $BiCh_2$-based compounds is induced by electron carrier doping into conduction band, which mainly consists of Bi 6p and Ch p orbitals [7,15]. In addition, recent studies have demonstrated that the emergence of bulk superconductivity requires optimization of crystal structure. This can be qualitatively described in terms of sufficient overlapping of Bi 6p and Ch p orbitals, namely, in-plane chemical pressure [16]. A typical route to induce in-plane chemical pressure is Se substitution for S site [17−19]. It has been reported that such pressure effects successfully suppressed the in-plane disorder of $BiCh_2$ conducting layer, which was revealed by means of synchrotron X-ray diffraction, extended X-ray absorption fine structure, and neutron diffraction [16,20−23]. Intrinsic superconductivity phase diagram mitigating in-plane disorder was recently established for $LaO_{1-x}F_xBiSSe$ [23].

On the carrier doping method for $BiCh_2$-based compounds, aliovalent ion substitution including partial substitution of $F^-$ for $O^{2-}$ ions in $REOBiCh_2$ (RE: rare earth or Bi) [24−30], or $RE^{3+}$ substitution for $A^{2+}$ ions in $AFBiCh_2$ (A: Ca, Sr, Eu) [31−36], have been typically employed to induce

(a) E-mail: xyxyxyxyx@xyxy.com
(b) Present address: Author's Instute, Author's University - Street and number, Postal Code City, Country.

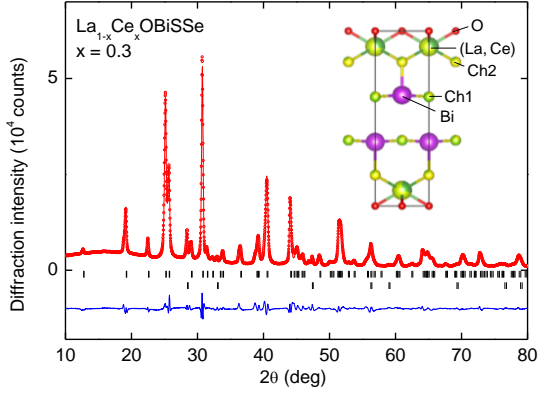

Fig. 1: (Color online) X-ray diffraction (XRD) pattern and the results of Rietveld refinement for La$_{1-x}$Ce$_x$OBiSSe with $x$ = 0.3. The circles and solid curve represent the observed and calculated patterns, respectively, and the difference between the two is shown at the bottom. The vertical marks indicate the Bragg diffraction positions for La$_{1-x}$Ce$_x$OBiSSe (upper) and CeO$_2$ (lower). The inset shows the schematic representation of crystal structure for $x$ = 0.3. Ch1 and Ch2 denote the in-plane and out-of-plane chalcogen sites, respectively. XRD patterns of $x$ = 0–0.9 are shown in Fig. S1 in Supporting Information.

superconductivity. It was also reported that the superconductivity appears via $M^{4+}$ substitution at La$^{3+}$ site in LaOBiS$_2$ (M: Ti, Zr, Hf, Th) [37]. Recently, zero resistivity was reported in CeOBiS$_2$, which is most likely due to electron doping via mixed valence state of Ce ions [38]. However, bulk superconductivity is yet to be confirmed in neither $M^{4+}$-doped LaOBiS$_2$ nor CeOBiS$_2$, on the basis of magnetic susceptibility or specific heat, which is a sensitive probe for bulk superconductivity.

In the present study, we synthesized a new BiCh$_2$-based superconductor system La$_{1-x}$Ce$_x$OBiSSe ($x$ = 0–0.9) and investigated the structural, electronic, and magnetic properties. Metallic conductivity was observed for $x \geq 0.1$, indicating electron carrier was doped by mixed valence state of Ce ions, as revealed by bond valence sum calculation and magnetization measurements. Bulk superconductivity, confirmed by zero resistivity and clear diamagnetic signal, was observed for $x$ = 0.2–0.6. To the best of our knowledge, this is the first report on the emergence of bulk superconductivity and metallic conductivity by substituting the RE site of REOBiCh$_2$ and using the mixed valence state of Ce ions in BiCh$_2$-based layered compounds.

**EXPERIMENTAL DETAILS** – Polycrystalline samples of La$_{1-x}$Ce$_x$OBiSSe with $x$ = 0, 0.1, 0.2, 0.3, 0.4, 0.5, 0.6, 0.7, 0.8, and 0.9 were prepared by a solid-state reaction method. Powders of CeO$_2$ (99.99%), La$_2$O$_3$ (99.9%), La$_2$S$_3$ (99.9%), Ce$_2$S$_3$ (99.9%), Bi (99.999%), S (99.99%), and Se (99.999%) were mixed in the stoichiometric ratio, pressed into pellet, and heated at 700 °C for 20 h in a sealed quartz tube. The obtained sample was ground, mixed, pelletized, and heated with the same heating condition.

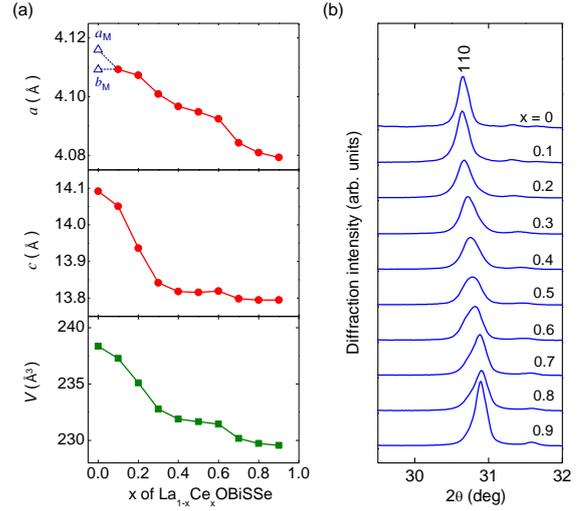

Fig. 2: (Color online) (a) Lattice parameters ($a$ and $c$) and lattice volume ($V$) of La$_{1-x}$Ce$_x$OBiSSe. For $x$ = 0, lattice parameters in monoclinic phase, which were calculated using synchrotron X-ray diffraction, was taken from literature [23]. (b) Expanded profiles of X-ray diffraction of 110 peak in tetragonal phase.

The obtained samples were characterized using powder X-ray diffraction (XRD) with a Cu K$\alpha$ radiation (Rigaku, Miniflex 600 equipped with a D/teX Ultra detector) by the $\theta$–$2\theta$ method. The crystal structure parameters were refined using the Rietveld method using RIETAN-FP code [39]. Crystal structure was depicted using VESTA software [40]. Chemical composition analysis was performed using scanning electron microscope (Hitachi, TM3030) equipped with an energy dispersive X-ray spectrometer (EDX; Oxford, SwiftED3000). Seebeck coefficient ($S$) at room temperature was measured by giving a temperature difference ($\Delta T$) of ~4 K, where the actual temperatures of both sides of the bulk were monitored by two thermocouples. The thermo-electromotive force ($\Delta V$) and $\Delta T$ were simultaneously measured, and the $S$ were obtained from the linear slope of the $\Delta V$–$\Delta T$ plots. Temperature ($T$) dependence of electrical resistivity ($\rho$) was measured using a four-terminal method. Magnetization as a function of $T$ was measured using a superconducting quantum interference devise (SQUID) magnetometer with an applied field of 10 Oe after both zero-field cooling (ZFC) and field cooling (FC) using Magnetic Property Measurement System (Quantum design, MPMS-3).

**RESULTS** – Figure 1 shows XRD pattern and the Rietveld fitting result for $x$ = 0.3 as a representative sample. Almost all the diffraction peaks can be assigned to those of the LaOBiSSe-type phase, indicating that this is a dominant phase in the samples, although minor diffraction peaks corresponding to CeO$_2$ impurity phase (< 6 wt%) were also observed. We have tested several annealing conditions for the Ce-substituted samples, but the CeO$_2$ impurity phase cannot be eliminated. Therefore, we have investigated the physical properties of the La$_{1-x}$Ce$_x$OBiSSe system with the samples obtained by the present annealing condition. Note that the sample with nominal $x = 1$ is excluded from further experiments because it contains significant amount of Bi$_2$Se$_3$ and CeO$_2$ impurities.

Table 1: Ce content (Ce/(La + Ce)) determined using chemical composition analysis, bond valence sum of RE site (BVS$_{RE}$), Curie constant (C), paramagnetic Curie−Weiss temperature ($\Theta_{CW}$), Ce valence and averaged La/Ce valence calculated using Curie−Weiss fitting for La$_{1-x}$Ce$_x$OBiSSe.

| $x$ | Ce/(La + Ce)[a] | BVS$_{RE}$ | $C$ (emu/K mol) | $\Theta_{CW}$ (K) | Ce valence[b] | Averaged La/Ce valence[b] |
|---|---|---|---|---|---|---|
| 0 | – | 2.98 | – | – | – | – |
| 0.1 | 0.10 | 3.04 | 0.011 | 1.4 | 3.64 | 3.07 |
| 0.2 | 0.20 | 2.78 | 0.029 | 1.0 | 3.58 | 3.12 |
| 0.3 | 0.30 | 3.02 | 0.066 | 0.4 | 3.47 | 3.14 |
| 0.4 | 0.36 | 3.09 | 0.107 | 0.1 | 3.36 | 3.12 |
| 0.5 | 0.49 | 3.10 | 0.153 | 0.2 | 3.37 | 3.19 |
| 0.6 | 0.62 | 3.11 | 0.194 | 0.0 | 3.38 | 3.23 |
| 0.7 | 0.70 | 3.16 | 0.241 | 0.2 | 3.34 | 3.24 |
| 0.8 | 0.80 | 3.14 | 0.271 | 0.2 | 3.35 | 3.28 |
| 0.9 | 0.91 | 3.15 | 0.334 | 0.4 | 3.32 | 3.29 |

[a] amount of Ce/(La + Ce) was determined using EDX analysis.
[b] Ce valence and averaged La/Ce valence were obtained using Curie−Weiss fitting of magnetization data measured from 5 to 20 K at 1 T (see text in detail).

Lattice parameters decrease with increasing $x$, as shown in Figure 2(a), primarily due to smaller ionic radius of Ce than that of La ions. Notably, it can be seen that diffraction peaks were broadened when the chemical composition was apart from the end member, as presented in Figure 2(b), most likely due to the local inhomogeneity of the crystal structure of these phases [41]. We note that symmetry lowering has been reported for $x = 0$ using synchrotron XRD equipped with a high-resolution one-dimensional semiconductor detector [23]. However, the previous report also determined that the crystal structure of the F-doped (electron-doped) LaOBiSSe system was tetragonal, without symmetry lowering. Therefore, it is most likely that the crystal structure of present Ce-doped (electron-doped) La$_{1-x}$Ce$_x$OBiSSe is also tetragonal $P4/nmm$. Chalcogen site occupancies were determined using Rietveld method imposing linear constraints to maintain their total occupancies in the stoichiometric one. Se occupancy at the in-plane Ch1 site was higher than 85% for all samples, indicating that Se ions selectively occupy the in-plane Ch1 site, rather than the out-of-plane Ch2 site, which is consistent with the observations in related BiCh$_2$-based layered compounds [17−19, 23]. We evaluated the valence state of La/Ce site using bond valence sum (BVS$_{\text{RE}}$). The BVS$_{\text{RE}}$ was calculated using the following parameters: $b_0 = 0.37$ Å, $R_0 = 2.172$ Å for La−O bond, 2.151 Å for Ce−O bond, 2.64 Å for La−S bond, 2.62 Å for Ce−S bond, 2.74 Å for La−Se bond, and 2.74 Å for Ce−Se bond [42,43]. Experimental bond distances between La/Ce and nine coordinating anions were determined using Rietveld analysis. Site occupancies of both La/Ce sites and chalcogen sites were included in the calculation of the BVS. The BVS$_{\text{RE}}$ of $x = 0$ was evaluated to be 2.98, corresponding to 3+ valence state of La ions. The BVS$_{\text{RE}}$ tends to increase with increasing $x$ and it reaches 3.15 for $x = 0.9$, as listed in Table 1. These results strongly suggest the mixed valence of Ce ions in La$_{1-x}$Ce$_x$OBiSSe, as in the case of CeOBiS$_2$, in which Ce mixed valence was demonstrated using X-ray absorption spectroscopy [44,45].

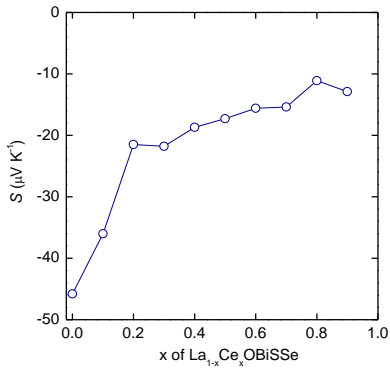

Fig. 3: (Color online) Room temperature Seebeck coefficient (S) for La$_{1-x}$Ce$_x$OBiSSe.

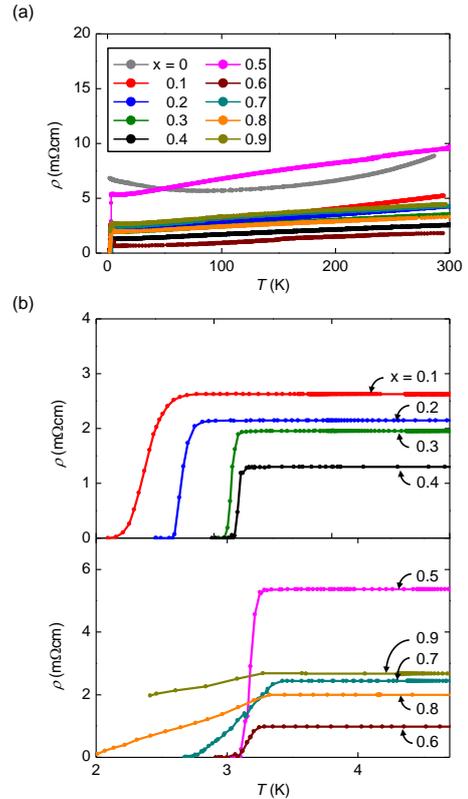

Fig. 4: (Color online) (a) Temperature (T) dependences of electrical resistivity ($\rho$) for La$_{1-x}$Ce$_x$OBiSSe. (b, c) $\rho$–$T$ plots below 4 K.

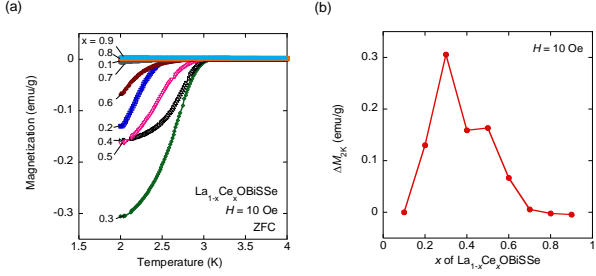

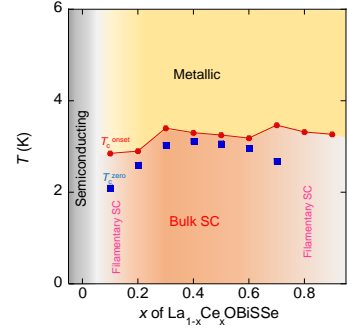

Fig. 5: (Color online) (a) Temperature dependences of magnetization with an applied magnetic field of 10 Oe for La$_{1-x}$Ce$_x$OBiSSe measured after zero-field cooling (ZFC). Magnetization measured after field cooling (FC) was shown in Fig. S2 in Supporting Information. (b) The amplitude of magnetization at 2 K ($\Delta M_{2\text{K}}$) as a function of $x$ for La$_{1-x}$Ce$_x$OBiSSe.

Fig. 6: (Color online) Superconductivity phase diagram of La$_{1-x}$Ce$_x$OBiSSe.

Figure 3 shows the $S$ at room temperature for $x = 0$–0.9. Negative values of $S$ indicate that the majority of carrier is electrons in all samples. Assuming single parabolic band model and acoustic phonon scattering as a dominant scattering process for carrier transport, the $S$ for metals or degenerate semiconductors is given by:

$$S = \frac{8\pi^2 k_B^2}{3eh^2} m^* T \left(\frac{\pi}{3n}\right)^{2/3} \quad (1)$$

where $n$ is the carrier concentration and $m^*$ is the effective mass of the carrier [46]. Absolute value of $S$ tends to decrease with increasing $x$, indicating that electron carrier was doped into the conduction band by Ce substitution.

Figure 4(a)−(c) shows the $T$ dependences of $\rho$ for $x = 0$–0.9. For $x = 0$, $\rho$ increases with decreasing $T$ below 70 K (Fig. 4(a)). In contrast, metallic behavior was observed for $x \geq 0.1$, namely, the $\rho$ decreases with decreasing $T$. The metallic conductivity originates from electrons that provided to BiCh$_2$ conducting layers [23] via the mixed valence of Ce ions. Regarding the superconducting characteristics, a steep decrease of $\rho$ attributable to $T_c^{\text{onset}}$ was observed for $x \geq 0.1$ (Figs. 4(b) and (c)). However, zero resistivity was obtained only for $x = 0.2$–0.6, and the highest $T_c^{\text{zero}}$ was 3.1 K for $x = 0.3$–0.5.

Figure 5(a) shows the temperature dependences of magnetization below 4 K with an applied magnetic field of 10 Oe after ZFC for La$_{1-x}$Ce$_x$OBiSSe. Clear diamagnetic signals corresponding to superconductivity are observed for $x = 0.2$–0.6, in consistent with the zero resistivity states of the samples in $\rho$–$T$ data, indicating that the observed superconducting states are bulk in nature. The largest shielding volume fraction and the $T_c$ of 3.1 K were obtained for $x = 0.3$, as plotted in Figure 5(b). The large shielding volume fraction exceeding the full diamagnetic response can be understood by the large demagnetization effect due to the presence of plate-like grains [47]. Magnetization measured under FC shows the onset to superconducting state as same as that measured under ZFC, as shown in Fig. S2, while diamagnetic signal is noticeably weak because the present samples are a type-2 superconductor with a strong pinning effect. Although measuring specific heat around $T_c$ is one of the best way to confirm the bulk nature of superconductivity for this system, it is challenging to characterize a superconducting transition from specific heat because magnetic entropy of Ce 4f electrons should overwhelm the electronic specific heat jump at a superconducting transition [48].

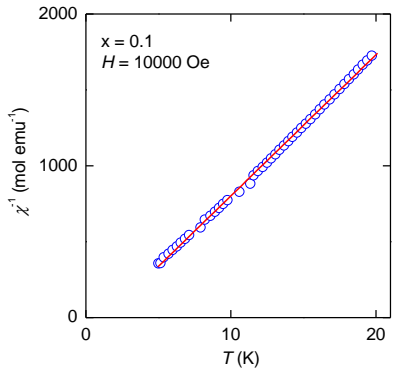

Fig. 7: (Color online) Temperature ($T$) dependences of reciprocal susceptibility ($\chi^{-1}$) of La$_{1-x}$Ce$_x$OBiSSe ($x = 0.1$) measured at 1 T. Red solid lines represent the Curie−Weiss fitting lines. Analogous data of other samples are shown in Fig. S3 in Supporting Information.

On the basis of the measurement results of electrical resistivity and magnetization, a superconductivity phase diagram was depicted in Figure 6. Weakly localized transport behavior was observed in $\rho$–$T$ for $x = 0$ below 70 K. This behavior was immediately suppressed, and metallic-like $\rho$–$T$ was observed for $x \geq 0.1$. The decrease of $\rho$ attributable to a superconducting transition was observed for $x = 0.1$–0.9, and $T_c^{onset}$ was ranged from 2.9 to 3.7 K in this region. While, zero resistivity and clear diamagnetic signal corresponding to the emergence of the bulk superconductivity states were achieved for $x = 0.2$–0.6. A dome-shaped superconductivity phase diagram, with the highest $T_c^{zero}$ of 3.1 K for $x = 0.3$–0.5, was obtained. Indeed, superconductivity of BiCh$_2$-based compounds are essentially sensitive to its local structure fluctuation. In the case of a Nd(O,F)BiS$_2$ single crystal, a bulk superconducting transition occurs below 5 K, while superconducting gap, measured by spectroscopy, remains up to 26 K [49]. Besides, a sharp superconducting transition was observed for optimally doped LaO$_{1-x}$F$_x$BiSSe with $x = 0.2$–0.5 [23]. This suggests that the present samples exhibiting broad phase transition ($x = 0.1, 0.7, 0.8, 0.9$) contains local inhomogeneity/fluctuations of their crystal structure, although these samples show relatively narrow peak widths of the XRD (Fig. 2(b)). To get further information about local structure inhomogeneity or structural fluctuations, experiments that sensitive to local structure such as extended X-ray absorption fine structure (EXAFS), are needed.

**DISCUSSION** – In order to examine the Ce valence state from magnetization measurements, Curie−Weiss fitting using $M/H = C/(T−\Theta_{CW})$ to the data of 1 T was performed, where $C$ is Curie constant and $\Theta_{CW}$ is paramagnetic Curie−Weiss temperature. Note that the magnetization data from 5 to 20 K are employed because $\chi^{-1}$ shows shoulder-like anomaly at around 100 K due to crystalline-electric-field effects [48]. Curie−Weiss fitting result of $x = 0.1$ is shown in Fig. 7 as a representative sample, and the analogous data of other samples are presented in Fig. S3. For $x = 0.1$, $C$ was evaluated to be 0.011 emu/K mol, indicating Ce valence state is +3.64. As listed in Table 1, Ce valence tends to decrease with increasing $x$, resulting in +3.32 for $x = 0.9$. The decrease of Ce valence seems to be correlated with increased in-plane chemical pressure because lattice constant $a$ decreases with $x$ (Fig. 2(b)). In the case of Eu$_{0.5}$La$_{0.5}$FBiS$_{2-x}$Se$_x$ system, the increase of in-plane chemical pressure suppress the mixed valence of Eu ions, which is demonstrated using a combined analysis of X-ray absorption near-edge structure and X-ray photoelectron spectroscopy [50]. It should be noted that while Ce valence tends to decreases with increasing $x$, the averaged RE valence (trivalent La and mixed-valent Ce) increase with $x$, in consistent with that obtained using BVS calculation (Table 1).

It is interesting to note that an anomaly in the temperature dependence of magnetization was observed at around 8 K for $x = 0.5$, as depicted in Fig. S4. It has been reported that magnetic moment of Ce$^{3+}$ ions of CeO$_{0.3}$F$_{0.7}$BiS$_2$ orders ferromagnetically, which was demonstrated using neutron scattering measurements [51]. Therefore, the magnetic anomaly for $x = 0.5$ at 8 K would be attributable to the ferromagnetic ordering of Ce$^{3+}$ moment, as well. The valence states and magnetism of Ce in La$_{1-x}$Ce$_x$OBiSSe are believed to be essentially sensitive to its chemical composition and/or (local) crystal structure. As an example, X-ray absorption spectroscopy revealed that mixed valence of Ce ions was suppressed by F doping for CeO$_{1-x}$F$_x$BiS$_2$ [44,45]. We note that the present samples are not single phase but contain CeO$_2$ impurity phase (< 6 wt%) (Figs 1 and S1). It has been reported that CeO$_2$ clusters may significantly contribute to the observed magnetic behavior [52]. Further investigation of local crystal/electronic structure is required to clarify the detailed relationship between magnetism and superconductivity in La$_{1-x}$Ce$_x$OBiSSe.

The highest $T_c$ of 3.1 K in La$_{1-x}$Ce$_x$OBiSSe is slightly lower than that of F-doped system, LaO$_{1-x}$F$_x$BiSSe ($T_c$ = 3.7 K) [14,23]. The present results suggest that substitution at different sites (substituting whether the O site or the La site) in blocking layer affect the superconducting properties of BiCh$_2$-based superconductors. For example, it has been reported that carrier doping via partial substitution for the site spatially far from the superconducting Fe plane results in higher $T_c$ in Fe-based superconductor BaFe$_2$As$_2$ [53]. It was also demonstrated that the decrease in $T_c$ originates from magnetic pair breaking by interaction of the localized 4$f$ orbitals in the RE dopants with the itinerant Fe 3$d$ orbitals for the corresponding compound [54].

**CONCLUDING REMARKS** – In summary, we have synthesized a new BiCh$_2$-based superconductor system La$_{1-x}$Ce$_x$OBiSSe. Polycrystalline samples for $x = 0$–0.9 were prepared by a solid-state reaction. Crystal structure analysis showed that both lattice constants $a$ and $c$ decrease with increasing $x$. BVS calculation and magnetization measurements showed mixed valence states of Ce ions in these compounds. Carrier doping into BiCh$_2$ conducting layer by Ce substitution is confirmed by the measurements of *S*. Metallic-like $\rho$–$T$ plots were observed for $x \geq 0.1$. Zero resistivity and clear diamagnetic signal corresponding to bulk superconductivity were obtained for $x = 0.2$–$0.6$. A dome-shaped superconductivity phase diagram, with the highest $T_c$ of 3.1 K for $x = 0.3$–$0.5$, was established. This work clearly showed the emergence of bulk superconductivity and metallic conductivity via mixed valence state of Ce ions in BiCh$_2$-based layered compounds.

\*\*\*

We thank OSUKE MIURA of Tokyo Metropolitan University for experimental support. This work was partly supported by Grant-in-Aid for Scientific Research (Nos. 15H05886, 16H04493, 17K19058, and 16K17944), JST-PRESTO (No. JPMJPR16R1), and JST-CREST (No. JPMJCR16Q6), Japan.

Supporting Information for

"Superconductivity in La$_{1-x}$Ce$_x$OBiSSe: carrier doping by mixed valence of Ce ions"

R. Sogabe[1], Y. Goto[1,*], A. Nishida[1], T. Katase[2,3], and Y. Mizuguchi[1]

1. Department of Physics, Tokyo Metropolitan University, Hachioji 192-0397, Japan
2. Laboratory for Materials and Structures, Institute of Innovative Research, Tokyo Institute of Technology, 4259 Nagatsuta, Midori, Yokohama, 226−8503, Japan
3. PRESTO, Japan Science and Technology Agency, 7 Gobancho, Chiyoda, Tokyo, 102-0076, Japan

E-mail: y_goto@tmu.ac.jp

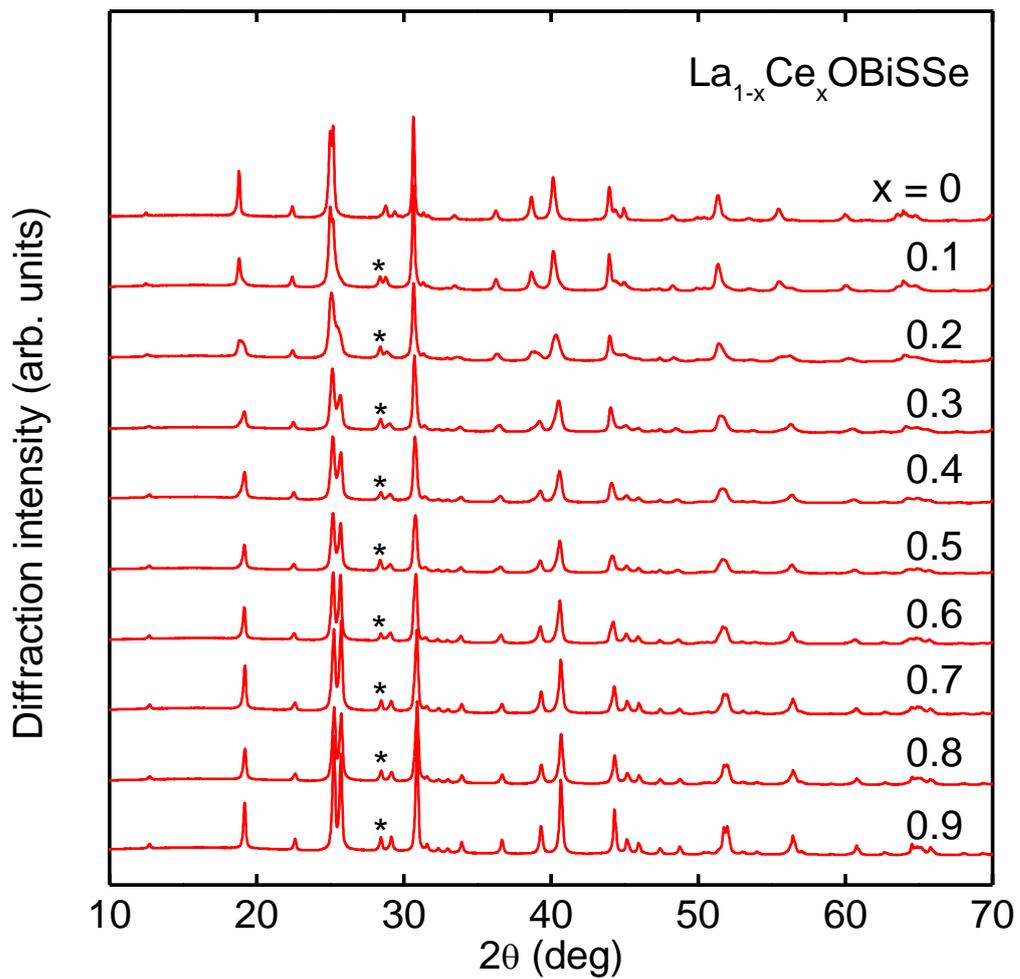

FIG. S1. XRD patterns for La$_{1-x}$Ce$_x$OBiSSe with $x$ = 0–0.9. Asteriks denote the diffraction peaks due to CeO$_2$ impurity phase.

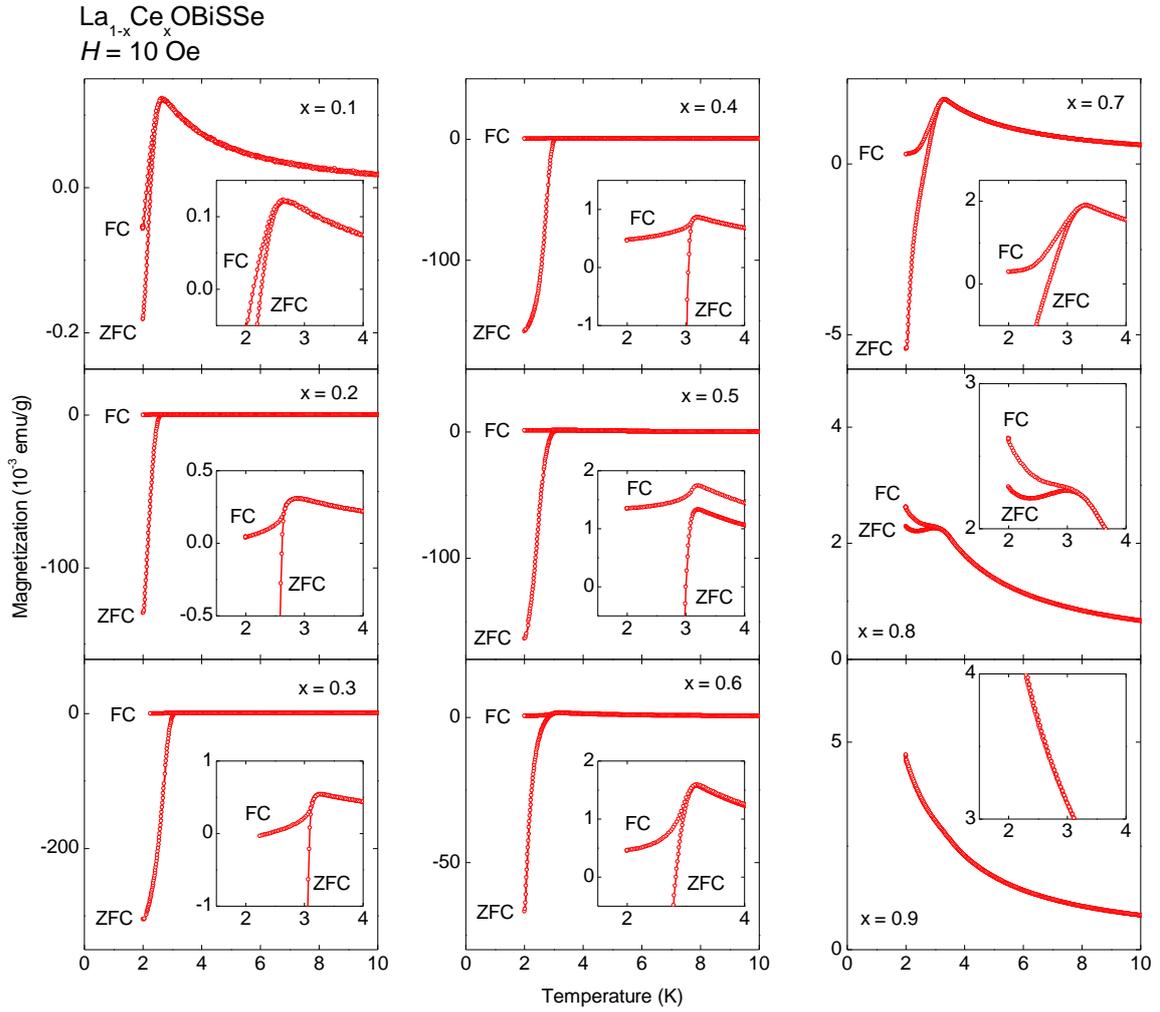

FIG. S2. (a) Temperature dependences of magnetization with an applied magnetic field of 10 Oe for La$_{1-x}$Ce$_x$OBiSSe measured after zero-field cooling (ZFC) and field cooling (FC). The inset shows expanded view near superconducting transition. For $x = 0.9$, magnetization measured after ZFC and FC nearly coincide.

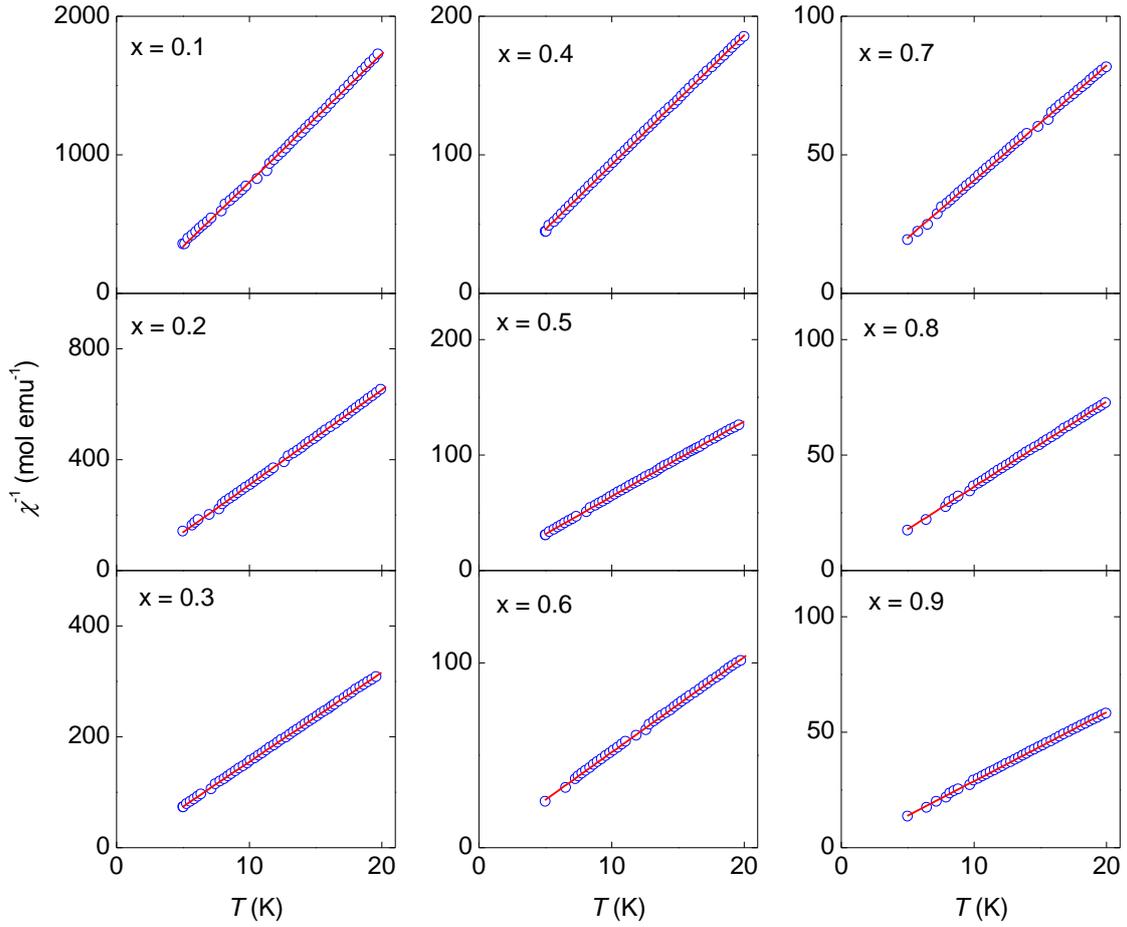

FIG. S3. Temperature ($T$) dependences of reciprocal susceptibility ($\chi^{-1}$) of La$_{1-x}$Ce$_x$OBiSSe measured at 1 T. Red solid lines represent the Curie–Weiss fitting lines. Note that the magnetization data from 5 to 20 K are employed because $\chi^{-1}$ shows shoulder-like anomaly at around 100 K due to crystalline-electric-field effects when magnetic field was applied along $c$-axis (R. Higashinaka *et al.* J. Phys. Soc. Jpn. **84**, 023702 (2015).).

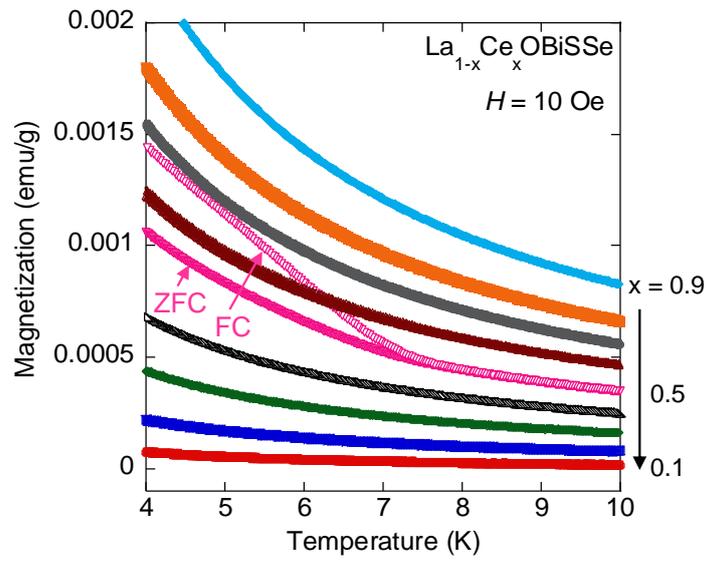

FIG. S4. Magnetization at 4−10 K for La$_{1-x}$Ce$_x$OBiSSe measured after ZFC and FC. Note that magnetic susceptibility after ZFC and FC nearly coincide except for $x$ = 0.5.